# KG-AST for Cheaper CI/CD: Budgeted Test Selection and AI Bug Review

2026-10-15 · Henry Yeung

**Author:** Henry Yeung, BSc (Mathematics, minor in Computer Science), The Chinese University of Hong Kong, henry930@gmail.com

## Abstract

Continuous integration and AI-assisted review spend the same budget on every commit, although few commits introduce bugs. We study KG-AST, which scores each commit's risk from git history and code-graph features before heavy tests or AI review run, so the budget goes to the riskiest commits first, and which hands the AI reviewer only a graph slice around the change. On 155 open-source projects in 11 languages, a single cross-language model trained on other projects ranks commits so that heavy testing of the riskiest 20% covers 49–64% of bug-introducing commits depending on the language, against 20% for random selection and 40–55% for testing the biggest change first. Every language beats the biggest-change baseline, by 3.8 to 13.2 points. In 20 Java projects the gain at a 20% budget is 15.1 points (64.6% vs 49.5%, 95% interval +11.6 to +18.4), and the model wins in 19 of the 20 projects. About twelve training projects suffice; richer features, tuning and cleaner labels did not help. An AI reviewer that sees only a commit's diff finds 26% of later-fixed defects exactly (32% with graph context), and about half of its findings that did not match a later fix were judged real by independent AI judges. A static call-graph selector runs about 8% of a test suite while including a test written for the fix in 85% of 67 fix commits. The Poincaré embedding used in our retrieval work adds nothing measurable here: commit-risk AUC, bug review and severity ranking are unchanged or worse. The labels are noisy and the budget metric assumes a heavy test catches the bug it covers; no CI minutes were measured.

## 1. Introduction

Testing and review are the most labour- and compute-intensive steps after code is written. Most pipelines treat every commit alike: the same suite runs, and the same reviewer or AI model reads the diff. Yet only a minority of commits introduce a bug (4% to 35% of commits per project in our data), so a budget spread evenly is largely spent on safe changes.

We ask a practical question: given a fixed budget of heavy tests or AI reviews, which commits should receive it? KG-AST answers with a risk score computed at commit time from the diff and from what git knew before the commit, and then limits the work done on a chosen commit to a slice of the code graph around the change. The companion paper (Yeung [12], in preparation) uses the same graph and a Poincaré embedding to cut the cost of sending code to an LLM; this paper concerns the CI/CD side.

Our question is cost, not completeness. A selector is useful if, at a small fraction of the budget, it covers most commits that will later be found to have introduced a bug. We therefore report coverage at a fixed budget, compared with sensible baselines that practitioners already use, rather than classification accuracy alone.

**Contributions.**

1. A commit-level risk model from git-only features that, at a 20% budget, covers 64.6% of bug-introducing commits in 20 Java projects (15.1 points above testing the biggest change first), with a leave-one-project-out protocol and bootstrap intervals.

2. Evidence that one cross-language model is enough: on 155 projects in 11 languages every language beats the biggest-change baseline, a model trained without seeing a language performs as well as one trained on it, and about twelve training projects saturate the result.
3. A measurement of what AI review of a selected commit finds and misses: 26–32% of later-fixed defects exactly, with a sampled audit of its apparent false positives.
4. A set of negative results: richer history features, tuning, cleaner labels, the Poincaré embedding, static AST detectors and graph-based severity ranking did not improve selection.

We do not measure CI minutes or money. The budget is the share of commits chosen for heavy work, and the claim holds as far as heavy work on a bug-introducing commit would expose its bug.

## 2. Background and Literature Review

This section summarises the concepts and methods the experiments rely on. Citations are listed in the References section with their web addresses.

### 2.1 Bug-introducing commits and the SZZ algorithm

A supervised commit-risk model needs to know which past commits introduced a bug. The SZZ algorithm (Śliwerski et al. [1]) finds them: identify a fix commit, take the lines it deleted or changed, and use `git blame` on the fix's parent to find the commits that last touched those lines. Those commits are labelled bug-inducing. We use `git blame -w -M` (ignoring whitespace and detecting moved lines), ignore blank, comment and import lines, and also label commits that were explicitly reverted.

SZZ labels are noisy. Herzig et al. [2] showed that mislabelled issue reports distort bug prediction, and Rosa et al. [3] built a developer-informed oracle for comparing SZZ implementations. Two biases matter for us. Bugs not yet found are labelled as clean, and large commits are more likely to be blamed because more lines can be traced. We guard against the second by also reporting a size-stratified AUC, and we repeat the main result with stricter labels (Section 5.3).

### 2.2 Change-level (just-in-time) defect prediction

Just-in-time models score a change when it is made, before the bug is known. Kamei et al. [4] predicted defect-inducing changes from change metrics grouped as size, diffusion, history, author experience and purpose. Nagappan and Ball [5] showed that relative code churn predicts defect density. Our 46 base features follow these groups (lines and files changed, entropy of the change, share of test code, content of added lines such as new branches and null checks, commit-message cues, author and file history), plus 13 features of defects known at commit time.

### 2.3 Test selection and budgeted evaluation

Predictive test selection learns from past results which tests are likely to fail for a change; Machalica et al. [6] report its use at Facebook. Commercial tools in this space sell risk scores and heat maps built from change history. Co-change mining (Zimmermann et al. [7]) is the history signal behind many of them. We evaluate with a budget view that matches how a team would use a score. With N test-period commits and a budget fraction b, the coverage at budget b is

$$R(b) = \frac{|\{\text{bug-inducing commits among the top } \lceil bN \rceil \text{ by score}\}|}{|\{\text{bug-inducing commits}\}|}$$

averaged over projects. A perfect ranking, a random ranking (R = b) and heuristics such as "largest change first" bracket the result. The metric assumes that a heavy test or review of a bug-inducing commit catches its bug, so it is an upper bound on detection.

### 2.4 Call-graph change impact

Static change impact analysis walks the call graph from the changed methods to the code, and the tests, that depend on them. We resolve calls by name without types, which is cheap and works on any snapshot but over- and under-approximates in the presence of overloading and dynamic dispatch.

### 2.5 LLMs for review and bug finding

LLMs are used for patch generation (SWE-bench, Jimenez et al. [8]) and for reviewing diffs. Evaluating review on public projects risks memorisation of fixes. Judging free-text findings needs a rubric, and an LLM judge (Zheng et al. [9]) is cheap but shows known biases, so we use independent judges, require a concrete triggering input for a finding to count as real, and report agreement.

### 2.6 Evaluation protocol

- **Splits.** Each project is split by time: the oldest 60% of commits train, the next 30% test, and the last 10% is dropped because bugs introduced there may not be fixed yet. Only fixes known at the training cutoff label training commits.
- **Cross-project generalisation.** Leave-one-project-out: a model for a project is trained on all other projects.
- **AUC and size-stratified AUC.** AUC ranks bug-inducing above clean commits; size-stratified AUC computes it within commit-size bins so size alone cannot explain it.
- **Model and uncertainty.** LightGBM gradient-boosted trees (Ke et al. [10]); bootstrap intervals (Efron [11]) resampling test commits.

## 3. System

KG-AST works per commit and needs no build of the customer's project. The computation that touches source code (graph, features, risk score) runs inside the customer's network; only a small graph slice is ever sent to an AI reviewer.

1. **Parse and score.** On each commit the system extracts the diff features and the file and author history known before the commit, and scores the commit with one cross-language gradient-boosted model. Parsing and scoring take seconds.
2. **Select.** Commits in a window are ranked by score. The top fraction b receives heavy work (full test suites, AI review, human review); the rest follow the cheap path.
3. **Slice.** For a selected commit the system builds the change radius: the changed methods, their callers and callees, and methods that historically changed with them (Paper 1 describes the Poincaré ranking of this set). A reverse walk of the call graph from the changed methods to test methods gives a targeted test set.
4. **Review and test.** An AI model reviews the diff with the slice as context and returns at most four findings with evidence; the selected tests run.
5. **Feed back.** Confirmed findings, dismissed findings and test failures are intended to flow back as labels and as heat on graph nodes, so that the model improves with use.

The table separates what is evaluated in this paper from what is a prototype or a design.

| Component | Status | Evaluated here |
|---|---|---|
| Graph builder (Java, Python, JS/TS; no build) | implemented | indirectly, through all results |
| Commit risk model (59 git features, one cross-language model) | implemented | Sections 5.1–5.3 |
| Call-graph test selector | experiment script | Section 5.5 (67 fix commits) |
| Change-radius slice | implemented (Paper 1) | Section 5.4 uses its context slices |
| AI review of a slice | experiment script | Section 5.4 (98 later-fixed bugs) |
| Heat-map dashboard (multi-source heat, high-risk list) | prototype | not evaluated |
| Environment build and test run, failure feedback | prototype: single run, manually started | not evaluated |
| Scheduled 24-hour test machine with AI triage of failures | design draft only | not evaluated |

## 4. Experimental Setup

The experiments fall into five groups, summarised below and detailed after the table.

| Group | Question | Data |
|---|---|---|
| Budgeted selection (Java) | How many bug-introducing commits does a risk-ranked budget cover, compared with simple heuristics? | 20 Java projects, most recent 3,000 commits each |
| Languages and training size | Does one model serve all languages, and how much training data does it need? | 155 projects in 11 languages; 30 Java projects for the learning curve |
| Label and feature checks | Do cleaner labels, more features or tuning change the result? | 26–30 Java projects |
| AI review of later-fixed bugs | What does an AI reviewer find, and how many apparent false positives are real? | 98 later-fixed bug commits and 42 clean commits from 15 Java projects |
| Static test selection | Can the call graph pick a small test set that includes the fix's test? | 67 fix commits in 7 Java projects |

**Projects and labels.** Projects are open-source repositories. Java projects: Apache Commons, Guava, Guice, Jackson, JavaParser, Jenkins, Joda-Time, jsoup, JUnit 5, Mockito, MyBatis, Netty, RxJava, Retrofit, Caffeine, Spring Framework and others (27 to 30 depending on the experiment). The 11-language set adds Python (14), JavaScript (19), TypeScript (12), PHP (12), Rust (14), C/C++ (11), Ruby (12), C# (10), Go (12) and Kotlin (10) projects; each contributes its most recent 2,500 commits, some from shallow clones of depth 5,000. Labels come from SZZ (Section 2.1). Fix commits are identified by message keywords (Conventional Commits prefixes for JavaScript and TypeScript), excluding documentation, refactoring, build and dependency changes.

**Features.** 46 base features (size and diffusion, entropy, test-code share, content of added and removed lines, message and time cues, author and file history) and 13 features of defects and activity known at commit time. Each language has its own file-extension, test-file and comment rules and regular expressions for the content features.

**Protocol.** Time split per project (60% train, 30% test, last 10% dropped), leave-one-project-out, LightGBM, macro average over projects, bootstrap 95% intervals over test commits. The reported number is coverage at budget b, the share of bug-introducing commits among the top b of commits by score.

**Baselines.** Random; biggest diff first; commit touching the most-changed file; commit touching the most-fixed file; a model using only change size; a model using only history; and a perfect-ranking oracle as an upper bound.

**AI review.** For each later-fixed bug, the bug-introducing commit is reviewed blind (the reviewer sees neither the fix nor the group) under four context slices with the same message, diff and six context methods: diff only; diff plus graph context; plus co-change history; plus Poincaré distance. Findings are scored by a second model against the real fix as exact, location or none. Cases are split into those where the fix changed methods the original commit never touched (56) and those where it changed only methods the commit touched (42). A sample of 36 findings that did not match a later fix was judged by two independent AI judges with access to the diff, today's code and later commits.

**Test selection.** At each of 67 fix commits that changed both code and tests, changed main methods are walked backwards through the name-resolved call graph (depth 1 to 3), optionally adding tests by the `FooTest` and `Foo` naming convention. The ground truth is the test methods the fix commit itself added or changed.

## 5. Results

**5.1 Budgeted selection in Java (20 projects).** Running heavy tests or review on the 20% of commits ranked riskiest by the git model covers 64.6% of bug-introducing commits, against 49.5% for the biggest-change heuristic and 19.6% at random. The gain over the biggest-change heuristic is 15.1 points at a 20% budget (95% interval +11.6 to +18.4), and the model wins in 19 of the 20 projects.

*[Figure not exported: Budget comparison, 20 Java open-source projects, leave-one-project-out; the perfect-ranking oracle (67.2% at 5%, 99.3% at 20%) is omitted from the plot]*

Across budgets the gain over the biggest-change heuristic is +10.5, +12.6, +15.1, +15.6 and +13.2 points at 5%, 10%, 20%, 30% and 50% (every lower bound above zero). A perfect ranking would cover 67.2% at a 5% budget and 99.3% at 20%, so there is room above the model. Size alone is strong: a model using only change size reaches 49.9% at 20%, and ranking by the biggest diff reaches 49.5%, so about two thirds of the model's coverage over random comes from size. The remaining 15 points come from change content, message cues and history. Budgets measured in changed lines rather than commits are not a sound cost model, because the cost of a test run rarely scales with diff size, and we do not use them.

**5.2 Eleven languages.** One model trained on the other 154 projects, whatever their language, beats the biggest-change heuristic in all 11 languages at a 20% budget. Coverage depends on the language, from 63.7% in Java to 49.2% in TypeScript, and part of this reflects the defect density and label quality of each ecosystem, not the model.

| Language | Projects | AUC | Coverage at 20%, model trained on other projects | Same, language never seen in training | Biggest change first | Gain over biggest change [95% interval] | Projects where model wins |
|---|---|---|---|---|---|---|---|
| Java | 29 | 0.843 | 63.7% | 64.2% | 50.5% | +13.2 [+9.4, +17.2] | 28 / 29 |
| Rust | 14 | 0.836 | 58.6% | 58.7% | 54.9% | +3.8 [+0.1, +7.9] | 10 / 14 |
| Python | 14 | 0.815 | 58.0% | 57.9% | 48.7% | +9.3 [+5.2, +13.9] | 12 / 14 |
| PHP | 12 | 0.823 | 57.8% | 58.3% | 47.8% | +10.0 [+3.3, +18.7] | 10 / 12 |
| C/C++ | 11 | 0.798 | 56.5% | 56.1% | 49.9% | +6.6 [+3.5, +9.6] | 9 / 11 |
| Ruby | 12 | 0.797 | 54.8% | 55.0% | 45.0% | +9.9 [+4.2, +15.7] | 9 / 12 |
| C# | 10 | 0.802 | 54.1% | 53.5% | 45.2% | +8.8 [+5.7, +12.1] | 10 / 10 |
| Go | 12 | 0.802 | 52.2% | 52.4% | 43.3% | +8.9 [+6.1, +12.2] | 12 / 12 |
| JavaScript | 19 | 0.794 | 50.7% | 49.7% | 40.2% | +10.6 [+6.6, +14.6] | 17 / 19 |
| Kotlin | 10 | 0.793 | 50.3% | 50.3% | 44.2% | +6.2 [+2.4, +9.9] | 7 / 10 |
| TypeScript | 12 | 0.798 | 49.2% | 48.8% | 40.1% | +9.1 [+5.0, +14.3] | 12 / 12 |

Random selection covers 20% in every language. Two findings stand out. First, a model that never saw a language performs as well as one trained on it, and training on all other projects beats training on the same language alone by 1 to 3 points, so a single model suffices and a new language needs only file-extension, test-file and regular-expression rules. Second, the earlier conclusion that JavaScript and TypeScript gain nothing was an artefact of sample size: with 8 projects the gain was +3.6 points [−4.0, +11.5], and with 31 projects it is +10.0 [+7.1, +13.2], with 29 of 31 projects improved. Rust gains least because the biggest-change heuristic is already strong there (54.9%). Some projects share an organisation (for example flask, click, jinja and werkzeug), which can slightly inflate leave-one-project-out results.

**5.3 What drives the score, and what does not help.** In 27 Java projects (about 71,000 commits) the feature groups contribute as follows. Size-stratified AUC is computed within eight commit-size bins, so it shows what remains after removing the effect of size.

| Features | AUC | Size-stratified AUC | Coverage at 10% | Coverage at 20% |
|---|---|---|---|---|
| Random | 0.504 | 0.506 | 11.9% | 20.7% |
| Commit size only | 0.767 | 0.590 | 32.9% | 52.4% |
| Size and diff features | 0.817 | 0.703 | 40.0% | 59.5% |
| + content of changes | 0.849 | 0.766 | 43.9% | 65.4% |
| + message and time | 0.852 | 0.771 | 45.3% | 64.5% |
| Author and file history only | 0.682 | 0.600 | 22.5% | 40.0% |
| All git features | 0.849 | 0.770 | 43.4% | 63.6% |

The strongest features by gain are the number of added `if` statements (22%), lines added (15%), the share of test lines (11%; commits with tests are safer), file age (8%) and the comment ratio (8%). A model trained only on a project's own history reaches AUC 0.81, below the 0.85 of a cross-project model, because one project has few positives; combining both is no better than the cross-project model.

With 30 Java projects (26 for the strict-label rows), these checks found no way to raise the result. The 20%-coverage figures are close to the 62.7% baseline in every case.

| Change | Coverage at 20% (30 Java projects) |
|---|---|
| Baseline: 46 features | 62.7% |
| + 13 known-defect-history features | 62.3% |
| + project-relative size features | 61.9% |
| + the test project's own earlier history in training | 62.9% |
| Six model settings (shallower, deeper, class-balanced, extra-trees, strong feature sampling) | 61.4% to 62.9% |
| Stricter labels (only fixes with an issue number or an explicit revert), trained and tested strict | 61.1% (AUC 0.817) |
| Looser training labels, strict test labels | 62.5% (AUC 0.822) |

The learning curve saturates early: coverage at 20% rises from 56.2% with 2 training projects to 60.8% with 8, 61.2% with 12 and 62.1% with all 29. A new customer therefore needs roughly ten public projects to match the plateau, not a large corpus. Further gains would need new information (CI results, logs, customer labels), not more git features or tuning.

**5.4 What AI review finds.** For 98 bug-introducing commits whose bug was fixed later, a blind AI reviewer found the bug exactly in 26% of cases from the diff alone and in 32% with graph context. At least one of the four context slices found it in 35% of cases, so about 65% are missed by all four. Exact-find rates with bootstrap intervals:

| Context slice | Fix touches methods the commit did not (n = 56) | Fix touches only methods the commit changed (n = 42) | All (n = 98) |
|---|---|---|---|
| Diff only | 0.23 [0.12, 0.36] | 0.29 [0.17, 0.43] | 0.26 [0.17, 0.35] |
| + graph context | 0.23 [0.12, 0.34] | 0.43 [0.29, 0.57] | 0.32 [0.22, 0.41] |
| + co-change history | 0.27 [0.16, 0.39] | 0.36 [0.21, 0.50] | 0.31 [0.21, 0.40] |
| + Poincaré distance | 0.29 [0.18, 0.41] | 0.31 [0.17, 0.45] | 0.30 [0.20, 0.39] |

Graph context helps overall (+0.061 [+0.010, +0.122]) and clearly when the fix stays among methods the commit touched (+0.143 [+0.048, +0.262]). It does not help when the fix lies in methods the commit never touched, which is the hard case, and history and Poincaré ranking do not beat the plain graph slice (Poincaré minus graph on the local cases is −0.119 [−0.238, −0.024]). Findings with confidence 0.7 or higher are rare: 60% of clean commits receive at least one finding, but only 0.02 per clean commit are at that confidence, so a high threshold gives few false alarms at the price of low recall (about 9%). A review costs roughly $0.04 (about $19 for 476 reviews).

Many findings that did not match a later fix may still be real. Two independent AI judges, agreeing on 34 of 36 cases, assessed a sample of 36 such findings, requiring a concrete triggering input.

| Confidence | Both judges: real defect | Both judges: false alarm | Disagree |
|---|---|---|---|
| 0.7 or higher (n = 12) | 7 (58%) | 2 | 3 |
| 0.5 to 0.7 (n = 24) | 12 (50%) | 11 | 1 |

Almost all of the real defects were already changed at the latest version of the code; only two persist today. Scaled to the unmatched findings, the precision of high-confidence findings rises from 60% (matching a later fix) to roughly 83%, and from 44% to 73% at confidence 0.5 or higher. This is an estimate: the sample is 36, the judges are AI models of the same family as the reviewer, no human has checked it, and the uncertainty is about 15 to 20 points.

**5.5 Static test selection from the call graph.** Running a small part of the suite chosen by the call graph includes a test written for the fix in most fix commits. This is an upper-bound style result, because the ground truth (tests the fix commit added or changed) is not independent of the code the fix touched.

| Selector (67 fix commits, median suite 1,250 test methods) | Share of suite run, mean / median | Fix's own test selected | Share of the fix's changed tests found |
|---|---|---|---|
| Call graph, depth 1 | 1.7% / 0.2% | 46.3% | 40.6% |
| Call graph, depth 2 | 6.5% / 0.9% | 61.2% | 54.0% |
| Naming convention only | 2.0% / 0.6% | 56.7% | 55.3% |
| Depth 2 + naming convention | 8.0% / 2.3% | 85.1% | 80.3% |
| Depth 3 + naming convention | 16.2% / 4.2% | 85.1% | 80.3% |
| Random selection of the same size | n/a | 23.2% | n/a |

Depth 2 plus the naming convention runs a median 2.3% of the suite (mean 8.0%) and includes a test written for the fix in 85% of commits, with no history needed. The limits are plain: unit tests only; only commits that changed tests (well-tested code); ground truth is "tests changed by the fix", not "tests that failed before the fix"; name-based call resolution; one pass of 67 commits. The Poincaré space was not used; the graph does the work.

**5.6 Negative results.** Several ideas that looked promising did not help.

| Idea | Setting | Result |
|---|---|---|
| Poincaré features for commit risk | Second stage on the git-only score, 10 Java projects, 5,139 test commits | AUC: baseline 0.856; + method complexity 0.848; + change heat 0.840; complexity and heat + Poincaré 0.849; the same + Euclid control 0.849. Poincaré and Euclid are indistinguishable and neither beats the baseline |
| Poincaré features for method-level bug prediction | Telegram Android, length-matched pairs (1,799 test pairs) | AUC: AST features 0.683; + Poincaré 0.685; + graph-hop control 0.676 (sampling error about 0.01) |
| Static AST detector across projects | Leave-one-project-out, 245 issue-linked fixes in 7 Java projects | Length-matched AUC 0.518 (logistic) and 0.551 (GBM); no better than method length |
| LLM picks defect-prone methods blind | 64 fix commits after the model's knowledge cutoff, 100-method pools | hit@10: Claude 9.4%, longest method 12.5%, random 4.7%; only 35.9% of fixes fall in the pool |
| Graph-based impact score to rank severity | 126 security fixes among 6,187 fixes in 16 Java projects | AUC 0.55 (change size 0.65); top 20% covers 23% of severe fixes against 20% at random |

The method-level result needs a caveat. The Telegram label (a small patch that adds a null, bounds or exception guard) is a proxy, and the strongest single signal there was the count of existing null checks, so the label may measure defensive coding as much as bugs. With real issue-linked fixes the detector did not generalise. Severity labels are message keywords such as CVE and denial of service, not true severity ratings, and the impact score is a crude name-based count of dependents.

## 6. Discussion and Limitations

**What the evidence supports.** A commit-risk score from git features alone lets a team spend heavy tests or AI review on a fifth of commits and still reach 50% to 64% of the bug-introducing ones, 4 to 15 points more than the usual heuristic of testing the biggest change. One model serves every language tested. The score is cheap, needs no build, and uses only information in the repository.

**What it does not support.**

- **Cost in money.** The budget is a share of commits, not minutes or dollars. A team whose CI environment is built once and shared across commits, and which runs full tests only on merge, saves less per skipped commit than one that pays per run. We have no customer CI data.
- **Detection.** Coverage assumes a heavy test or review of a bug-inducing commit catches its bug. AI review found 26–32% of later-fixed defects exactly, and about 65% were missed by all four context slices.
- **The graph and the Poincaré space.** For commit selection they add nothing measurable. The graph's value is in the slice around a selected commit (Paper 1) and in call-graph test selection, not in the risk score. We therefore do not claim hyperbolic geometry reduces CI cost.

**Limitations.**

1. Labels are SZZ labels and keyword-identified fixes: noisy, biased toward large commits, and blind to bugs not yet found. Stricter labels did not change the result, but none is verified by a human.
2. All projects are open source and widely used; models in the AI review may have seen their histories. Absolute numbers may be high; comparisons between context slices remain fair.
3. Several projects share an organisation, so leave-one-project-out slightly overstates generalisation.
4. Each project contributes its most recent 2,500 to 3,000 commits, and some come from shallow clones, so early history features may be truncated.
5. The AI audit uses AI judges and 36 samples. The test-selection ground truth is partly circular. Languages other than Java have fewer projects.
6. The test machine, scheduled testing, AI triage of failures and the feedback loop into heat are designs or prototypes and are not evaluated. We did not compare with commercial test-selection products on the same data, because their models cannot be reproduced.
7. There is no customer data, no real CI log and no measurement of developer time saved.

## 7. Conclusion and Future Work

A risk score computed from git features, with no build and one model for all languages, concentrates heavy testing and AI review where bugs are introduced: the riskiest fifth of commits covers 49% to 64% of bug-introducing commits across 155 projects in 11 languages, and 15 points more than the biggest-change heuristic in Java. The result is stable under changes of features, tuning and labels, and saturates with about twelve training projects. The code graph is useful after selection, for slicing context for an AI reviewer and for picking a small test set, but the Poincaré embedding adds nothing to the selection itself.

The next steps address the gaps that matter to a user.

1. Measure real cost: replay CI logs from a few projects or customers and report minutes and money saved at a given miss rate.
2. Add new information that git lacks: test and CI failures, logs and customer-confirmed labels, which our experiments suggest are the only route to further gains.
3. Replace SZZ labels with confirmed ones, and check the AI review sample by humans.
4. Close the loop: schedule tests by risk, triage failures by an AI model, and feed confirmed outcomes back as labels, then evaluate with mutation testing and replayed fix records instead of "changed again later".
5. Extend the call-graph test selector beyond Java unit tests to interface, API and flow tests.

## Code availability

The KG-AST source code is not publicly released. Researchers interested in checking or reproducing the results can contact the author at henry930@gmail.com; access to the code is decided case by case. The projects analysed are public open-source repositories.

## References

*Bracketed numbers in the text, such as [4], refer to the entries below in order.*